\documentclass[pra,aps,onecolumn,showpacs,superscriptaddress,10pt]{revtex4-2} 

\pdfoutput=1

\usepackage{amsmath,amssymb,graphicx}
\usepackage[colorlinks=true, allcolors=blue]{hyperref}
\usepackage{textcomp}

\begin{document}

\title{Cavity-enhanced sum-frequency generation of blue light with near-unity conversion efficiency}

\author{Hugo Kerdoncuff}
\affiliation{Danish Fundamental Metrology, Kogle Alle 5, DK-2970 H\o rsholm, Denmark}

\author{Jesper B. Christensen}
\affiliation{Danish Fundamental Metrology, Kogle Alle 5, DK-2970 H\o rsholm, Denmark}

\author{T\'ulio B. Brasil}
\affiliation{Niels Bohr Institute, University of Copenhagen, Blegdamsvej 17, 2100 Copenhagen, Denmark}
\affiliation{Instituto de F\'isica, Universidade de S\~ao Paulo, P. O. Box 6 6318, 05315-970 S\~ao Paulo, Brazil}

\author{Valeriy A. Novikov}
\affiliation{Niels Bohr Institute, University of Copenhagen, Blegdamsvej 17, 2100 Copenhagen, Denmark}

\author{Eugene S. Polzik}
\affiliation{Niels Bohr Institute, University of Copenhagen, Blegdamsvej 17, 2100 Copenhagen, Denmark}

\author{Jan Hald}
\affiliation{Danish Fundamental Metrology, Kogle Alle 5, DK-2970 H\o rsholm, Denmark}

\author{Mikael Lassen}\email{Corresponding author: ml@dfm.dk}
\affiliation{Danish Fundamental Metrology, Kogle Alle 5, DK-2970 H\o rsholm, Denmark}



\begin{abstract}
We report on double-resonant highly efficient sum-frequency generation in the blue range. The system consists of a 10-mm-long periodically poled KTP crystal placed in a double-resonant bow-tie cavity and pumped by a fiber laser at 1064.5~nm and a Ti:sapphire laser at 849.2~nm. An optical power of 375~mW at 472.4~nm in a TEM$_{00}$ mode was generated with pump powers of 250~mW at 849.2~nm and 200~mW at 1064.5~nm coupled into the double-resonant ring resonator with 88$\%$ mode-matching. The resulting internal conversion efficiency of 95($\pm 3$)$\%$ of the photons mode-matched to the cavity constitutes, to the best of our knowledge, the highest overall achieved quantum conversion efficiency using continuous-wave pumping. Very high conversion efficiency is rendered possible due to very low intracavity loss on the level of 0.3$\%$ and high nonlinear conversion coefficient up to 0.045(0.015) W$^{-1}$. Power stability measurements performed over one hour show a stability of 0.8$\%$. The generated blue light can be tuned within 5 nm around the center wavelength of 472.4 nm, limited by the phase-matching of our nonlinear crystal. This can however be expanded to cover the entire blue spectrum (420 nm to 510 nm) by proper choice of nonlinear crystals and pump lasers. Our experimental results agree very well with analytical and numerical simulations taking into account cavity impedance matching and depletion of the pump fields.
\end{abstract}


\maketitle


\section{Introduction}

Stable and reliable coherent light sources in the ultraviolet (UV) to blue wavelength range have become essential for a large variety of applications in fundamental research, medical diagnosis, and industrial processing \cite{Ye2008,Yun2017,Wang2017}. However, blue and UV light sources are relatively difficult to produce with high output powers and long lifetime, and regular laser diodes do not cover all wavelengths. Nonlinear frequency conversion techniques are therefore very attractive for the generation of blue coherent light covering the complete wavelength range \cite{cankaya2014highly,Moura2016}. Three-wave parametric interaction with second-order nonlinear crystals enables conversion to wavelengths that potentially cover the whole UV to visible region. The wavelength conversion bandwidth and efficiency of the second-order nonlinear interaction depends, among others, on the spatial and temporal overlap of the two fundamental beams and their phase matching with the third beam. Sum-frequency generation (SFG) is a parametric process, meaning that the photons satisfy energy conservation: $\omega_3=\omega_1+\omega_2$, where $\omega_3$ is the sum frequency (SF) and $\omega_{1,2}$ are the fundamental pump frequencies. The generation of visible light from optically-pumped solid-state and semiconductor lasers is usually achieved via second-harmonic generation (SHG), which is a special case of SFG, where $\omega_1=\omega_2$. However, a main limitation of SHG-based visible light sources is the availability of near-infrared laser lines, consequently the whole UV to blue wavelength range cannot be covered \cite{Polzik92,Ast2011, Liu2018,Cui2018}. SFG, on the other hand, allows the generation of light in the complete UV to visible range, by having one of the pump lasers chosen for wavelength selectability and tunability. The SF field potentially inherits all the properties of the fundamental pump lasers, such as beam quality, noise level and linewidth. A well-known and studied application of SFG is the generation of light at the Sodium D$_2$ resonance at 589 nm by using the 1064 nm and 1319 nm laser lines of YAG lasers \cite{Bienfang03,Mimoun2008}. Recently, SFG with high conversion efficiencies has shown huge potential for quantum frequency conversion, which bridges the gap between a variety of quantum systems by allowing frequency conversion of quantum states while preserving the quantum correlations \cite{Tanzilli2005, Allgaier2017}.

The SFG process can be operated in various optical configurations, such as single-pass \cite{Johansson2005, Hansen2015}, single-resonant \cite{Cheung94}, double-resonant (DR) \cite{Risk92, Kretschmann97, vance1998continuous}, or triple-resonant systems \cite{Zhang2001}. The single-pass SFG process is independent of the phase relation between the two fundamental pump fields, thus phase locking of the two lasers is not necessary and the system can be made compact and simple. However, high-efficiency nonlinear conversion with continuous-wave (CW) lasers in a single-pass system requires the use of very high optical input powers or of specially designed waveguide devices. Pulsed lasers with high peak powers are therefore preferred for reaching high efficiency in single-pass systems, whereas resonant cavities with strong buildup of circulating power are employed for increasing the conversion efficiency with CW pump fields. In single-resonant SFG, only one of the two fundamental pump fields is resonantly enhanced by the cavity. In double-resonant SFG (DR-SFG), the circulating powers of both pump fields are resonantly enhanced. Since the pump fields have different frequencies, this sets constrains to the design of the resonant cavity. However, very high quantum conversion efficiency (QCE), i.e. fraction of converted photons, close to unity and only limited by passive losses in the cavity and the nonlinearity, can be achieved in the DR configuration.

In this article, we present the experimental realization of highly efficient DR-SFG in an external resonator in which the two pump beams at 849.2~nm and 1064.5~nm are maintained at the resonance simultaneously by applying a double phase-locking scheme. We use a DR bow-tie cavity with an 8$\%$ input coupler for both wavelengths. The input coupler is optimized with respect to the available amount of optical power at 849.2~nm. Blue light in a TEM$_{00}$ mode with an optical power of 375 mW at 472.4 nm is generated with 88$\%$ of the 250 mW of 849.2~nm power and 200 mW of 1064.5~nm power coupled into the DR bow-tie ring cavity. This demonstrates an internal QCE of 95($\pm 3$)$\%$ of the photons mode-matched to the cavity. To the best of our knowledge it is the highest overall achieved QCE for SFG with CW pumping \cite{Cui2018}. The external QCE of 84($\pm 3$)$\%$ of the incident photons is thus essentially limited by the mode-matching to the cavity. Our experimental results agree very well with our analytical model and numerical simulations.

\section{Theory of sum-frequency generation}

To describe SFG in a DR cavity, we first quickly review the basic theory of single-pass SFG in the plane-wave approximation. Consider pump fields with wavelengths $\lambda_1$ and $\lambda_2$ and input pump powers $P_{1,\mathrm{in}}$ and $P_{2,\mathrm{in}}$, assumed, without loss of generality, to obey $P_{1,\mathrm{in}} \lambda_1 > P_{2,\mathrm{in}} \lambda_2$. By defining $m = P_{2,\mathrm{in}} \lambda_2 /P_{1,\mathrm{in}} \lambda_1 < 1$,  the SF output power at the wavelength  $\lambda_3 = (\lambda_1^{-1} + \lambda_2^{-1})^{-1}$ is given by \cite{moore2002resonant}
\begin{equation}
    P_{3,\mathrm{out}}= (\lambda_2 / \lambda_3)  P_{2,\mathrm{in}}  \mathrm{sn}^2 \left\{ \left.\left[(\lambda_3 / \lambda_2) \alpha P_{1,\mathrm{in}}\right]^{1/2} ~ \right| m \right\}.
    \label{eq:P3full}
\end{equation}
The solution given in terms of the Jacobian elliptic sine highlights the periodic nature of the SFG process with a maximum SF power given by $(\lambda_2/\lambda_3) P_{2,\mathrm{in}}$. The parameter $\alpha$ is the nonlinear conversion coefficient often quoted in literature and given in units of inverse watts. Its value depends on the specifications of the nonlinear crystal and on the overlap and focusing of the pump modes. Expansion of Eq.~(\ref{eq:P3full}) to first order in $\alpha$, leads to the common approximate expression with the SF power being linear with respect to the pump powers \cite{boyd2003nonlinear,kaneda1997theoretical}
\begin{equation}
    P_{3,\mathrm{out}}^{(\mathrm{app.})} = \alpha P_{1,\mathrm{in}} P_{2,\mathrm{in}}.
    \label{eq:P3approx}
\end{equation}
This constitutes the undepleted-pump approximation where it is assumed that the pump fields remain undepleted throughout the nonlinear interaction in the crystal. The approximation is often valid due to the low values of nonlinear conversion coefficients $\alpha$, e.g. on the order of $10^{-2}$-$10^{-3}$~W$^{-1}$ with KTP crystals.

As the SF power increases, energy conservation dictates that the pump powers decrease. The output pump powers are given by the relations \cite{moore2002resonant,Armstrong1962}
\begin{subequations}
\begin{equation}
P_{1,\mathrm{out}} = \mathrm{dn}^2 \left\{ \left.\left[(\lambda_3 / \lambda_2)\alpha P_{1,\mathrm{in}}\right]^{1/2} ~ \right| m \right\}  P_{1,\mathrm{in}} \equiv (1- \Gamma_1) P_{1,\mathrm{in}}, 
\label{eq:P1_infull}
\end{equation}
\begin{equation}
P_{2,\mathrm{out}} = \mathrm{cn}^2 \left\{ \left.\left[ (\lambda_3 / \lambda_2)\alpha P_{1,\mathrm{in}} \right]^{1/2} ~ \right| m \right\}  P_{2,\mathrm{in}} \equiv (1-\Gamma_2) P_{2,\mathrm{in}},
\label{eq:P2_infull}
\end{equation}
\label{eq:P1andP2}%
\end{subequations}
where the defined conversion parameters $\Gamma_1$ and $\Gamma_2$ are functions of both of the input pump powers. As for the SF power, expansion of the Jacobian elliptic functions to first order in the nonlinear conversion coefficient gives the approximate expressions \cite{kaneda1997theoretical}
\begin{subequations}
\begin{equation}
P_{1,\mathrm{out}}^{(\mathrm{app.})} = \left[ 1 - (\lambda_3 / \lambda_1) \alpha P_{2,\mathrm{\mathrm{in}}}  \right] P_{1,\mathrm{in}} \equiv (1-\Gamma_1^{(\mathrm{app.})}) P_{1,\mathrm{in}}, 
\label{eq:P1_inapprox}
\end{equation}
\begin{equation}
P_{2,\mathrm{out}}^{(\mathrm{app.})} = \left[ 1 - (\lambda_3 / \lambda_2) \alpha P_{1,\mathrm{in}}  \right] P_{2,\mathrm{in}} \equiv (1- \Gamma_2^{(\mathrm{app.})}) P_{2,\mathrm{in}},
\label{eq:P2_inapprox}
\end{equation}
\label{eq:P1andP2approx}%
\end{subequations}
where we used the formulas $\mathrm{dn} (u|m) \approx 1 -  u^2 m/2$ and $\mathrm{cn} (u|m) \approx 1 -  u^2/2$. Both Eqs.~(\ref{eq:P3full}) and (\ref{eq:P1andP2}) as well as the approximate expressions Eqs.~(\ref{eq:P3approx}) and (\ref{eq:P1andP2approx}) conserve the total initial power $P_{1,\mathrm{in}} + P_{2,\mathrm{in}} $.  

Having established expressions for the single-pass SFG configuration, we move on to the DR case. The SF field does not experience a cavity effect, and therefore the generated SF power may be found using Eq.~(\ref{eq:P3full}) while replacing the input powers $P_{1,\mathrm{in}} $ and $P_{2,\mathrm{in}}$ by the circulating pump powers $P_{1,\mathrm{circ}} $ and $P_{2,\mathrm{circ}}$. These circulating powers are enhanced with respect to the pump powers incident on the cavity, $P_{1,\mathrm{inc}}$ and $P_{2,\mathrm{inc}}$,  according to \cite{Ismail16}
\begin{subequations}
\begin{equation}
P_{1,\mathrm{circ}} = \frac{1-R_{1,\mathrm{in}}}{ \left[ 1 - \sqrt{R_{1,\mathrm{in}} (1-\Delta_1)(1- \Gamma_1) } \right] ^2} \ P_{1,\mathrm{inc}} ,
\label{eq:P1_v2}
\end{equation}
\begin{equation}
P_{2,\mathrm{circ}} =\frac{1-R_{2,\mathrm{in}} }{\left[ 1 - \sqrt{ R_{2,\mathrm{in}} (1-\Delta_2)(1- \Gamma_2) } \right]^{2}} \ P_{2,\mathrm{inc}} , 
\label{eq:P2_v2}
\end{equation}
\label{eq:P1andP2_v2}%
\end{subequations}
where $R_{i,\mathrm{in}}$ is the reflectivity of the cavity input mirror, $\Delta_i$ is the passive transmission loss in the cavity (excluding the input mirror), and $\Gamma_i$ is the SFG-induced conversion defined in Eqs.~(\ref{eq:P1andP2}), again with $P_{1,\mathrm{in}} $ and $P_{2,\mathrm{in}}$ replaced by $P_{1,\mathrm{circ}} $ and $P_{2,\mathrm{circ}}$. Due to the nonlinear coupling between the circulating pump powers, Eqs.~(\ref{eq:P1andP2_v2}) constitutes two simultaneous nonlinear equations, which, in general, must be solved by numerical means to find the circulating powers, and, thus, the generated SF power. 

\begin{figure}[h!]
\centering
\includegraphics[scale=0.65]{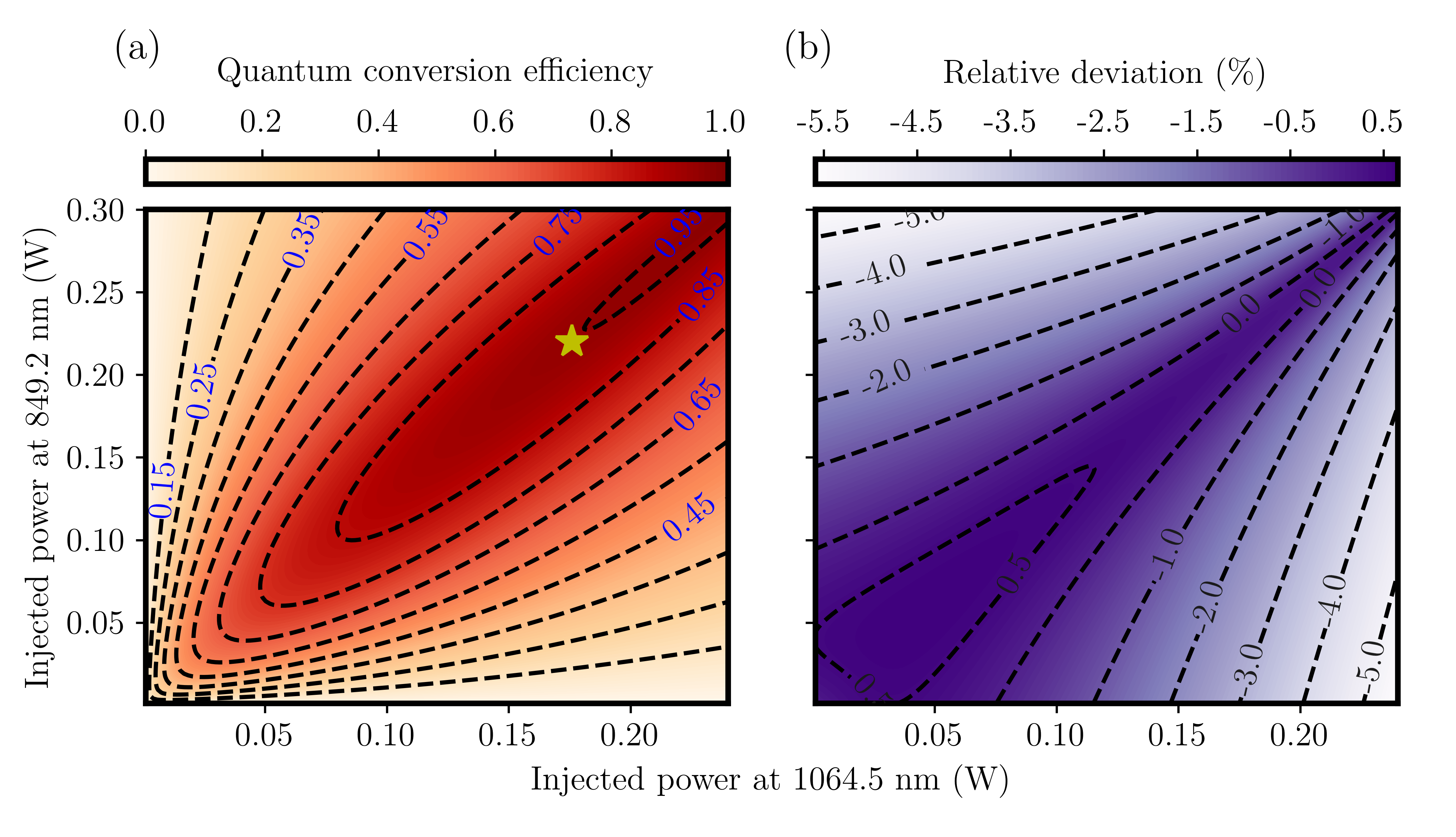}
\caption{(a) Total QCE as a function of the incident pump powers. (b) Calculated relative deviation between QCEs obtained by the perturbative and the rigorous theory. Parameters used in the simulations are $R_{1,\mathrm{in}} = R_{2,\mathrm{in}} = 0.92$, $\Delta _1 = \Delta_2= 0.003$, and $\alpha = 0.045~\mathrm{W}^{-1}$.  }
\label{Fig:Theory}
\end{figure}

To assess the efficiency of the SFG process, we define the total QCE, $\eta$, as the harmonic mean of the two individual QCEs:
\begin{equation}
\eta = \frac{2 \lambda_3 P_{3,\mathrm{out}}}{\lambda_1 P_{1,\mathrm{inc}} + \lambda_2 P_{2,\mathrm{inc}} } = \frac{2}{\eta_1^{-1} + \eta_2^{-1}},
\label{eq:QCE}%
\end{equation}
where $\eta_i = \lambda_3 P_{3,\mathrm{out}} /\lambda_i P_{i,\mathrm{inc}}$. Defined like this, $\eta$ quantifies the fraction of converted photons to all incident pump photons, and can therefore only approach unity in case of similar incident pump photon fluxes. Fig.~\ref{Fig:Theory}(a) shows the simulated QCE as a function of the two incident pump powers using parameters alike those reported in the experimental section below ($R_{1,\mathrm{in}} = R_{2,\mathrm{in}} = 0.92$, $\Delta _1 = \Delta_2= 0.003$, and $\alpha = 0.045~\mathrm{W}^{-1}$). The star marker indicates the experimentally achieved result of 95$\%$ total QCE. Such a high QCE is attainable due to ultra-low intracavity losses and a highly balanced cavity configuration that enables both pump fields to be nearly impedance matched to the cavity simultaneously. Fig.~\ref{Fig:Theory}(a) was obtained using the rigorous theory of Eqs.~(\ref{eq:P3full}) and (\ref{eq:P1andP2}).

There has been some debate over the validity of the approximations in Eqs.~(\ref{eq:P3approx}) and (\ref{eq:P1andP2approx}) in relation to SFG in a DR configuration \cite{moore2002resonant,Mimoun2008}. To investigate the error made in using the approximate forms, Fig.~\ref{Fig:Theory}(b) shows the relative error in QCE, $(\eta^{(\mathrm{app.})} - \eta )/\eta $, for the same experimental parameters as considered in Fig.~\ref{Fig:Theory}(a). The approximate theory agrees well (within $1~\%$) with the rigorous theory as long as the pump fluxes are roughly balanced. This even applies at high conversion efficiencies as long as the incident powers remain a small fraction of the circulating powers \cite{Mimoun2008}. However, as the incident fluxes become imbalanced, which is often what has been considered experimentally \cite{Mimoun2008, Lu2009, Liu2018}, we find that the approximate theory begins underestimating the QCE. The reason behind this is that the approximate theory overestimates the nonlinear loss experienced by the weaker pump. As a result of the increased nonlinear loss, the circulating power decreases, as does the calculated QCE.


\section{Experimental setup}

\begin{figure}[h!]
\centering\includegraphics[width=12cm]{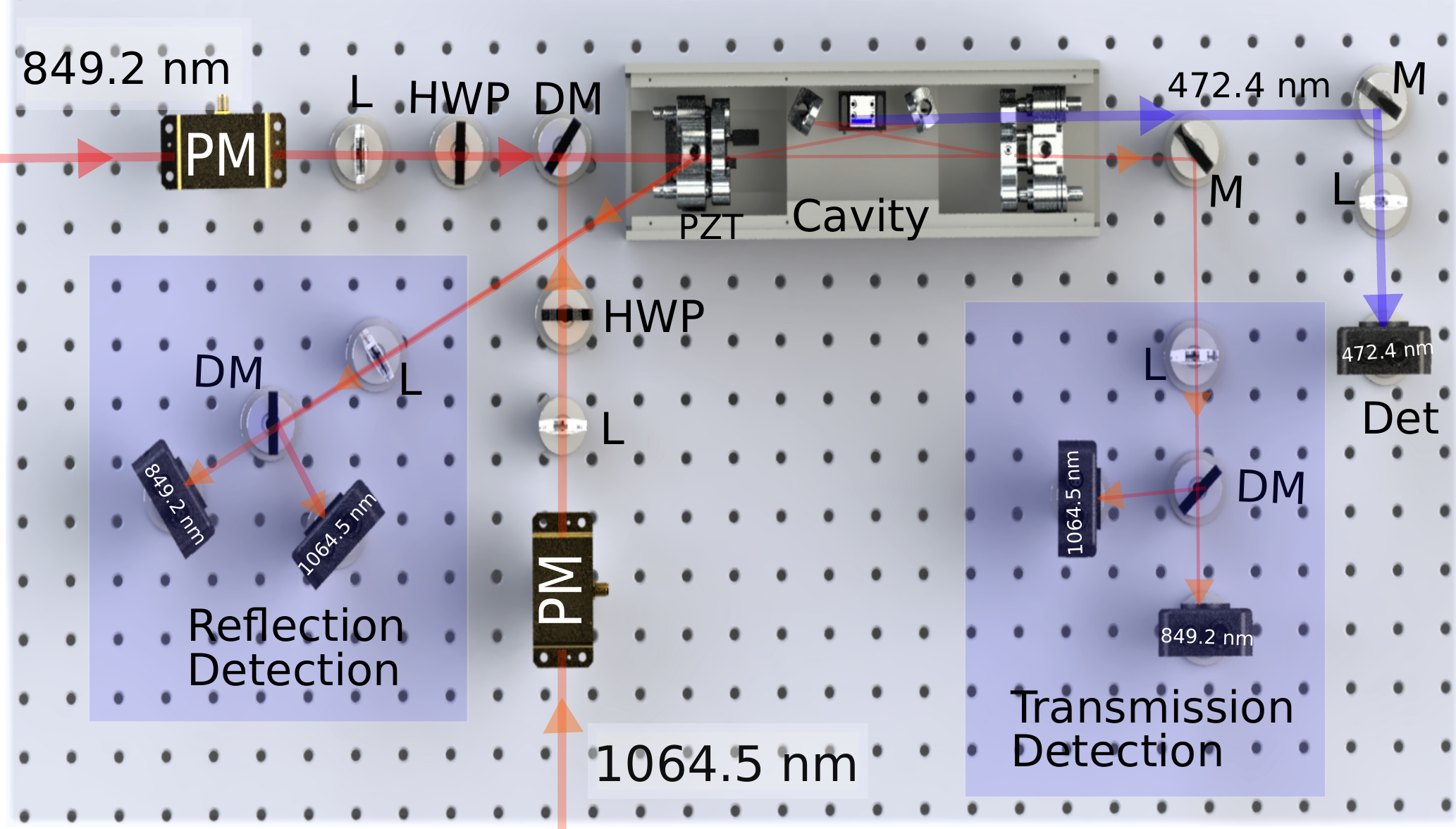}
\caption{Block diagram of the experimental setup with the main components. The reflection and the transmission from the cavity is measured with four 150 MHz bandwidth photodetectors. DM: dichroic mirrors, PM: phase modulators, HWP: half-wave plate, L:~lenses, PZT: Piezo-electrical element, M: mirrors. }
\label{Fig:setup}
\end{figure}

The schematic of our setup is depicted in Fig.~\ref{Fig:setup}. The pump sources used for the SFG are a 1064.5~nm fiber laser with 10~kHz linewidth and up to 10~W output power, and a Ti:sapphire laser tunable in the range from 770~nm to 880~nm with 75~kHz linewidth and delivering a maximum power of 250~mW to the cavity. We use a bow-tie configuration for the DR cavity in order to avoid destructive interference that may result from a double-pass linear cavity. Furthermore, the circulating beam encounters the passive loss in the crystal only once per round trip. The nonlinear crystal is a type-0 periodically poled KTP (PPKTP) crystal (Raicol Crystals Ltd.) with dimensions of (1$\times$ 1 $\times$ 10) mm$^3$ and a poling period of 6.1 \textmu m. PPKTP was choice due its high nonlinear coefficient and is a very well-known crystal for the generation of non-classical light and has been intensively investigated \cite{Lassen2007}. Its facets are anti-reflection coated for 1064 nm and 852 nm. The crystal is positioned on a copper platform mounted on a thermoelectric element and its temperature is monitored with a thermistor for temperature control. Controlling the temperature of the nonlinear crystal allows tuning the refractive index of the crystal in order to optimize phase-matching of the fundamental pump and SF fields. Our bow-tie cavity consists of two plane mirrors and two curved mirrors with a radius of curvature of 38 mm. All mirrors are super-polished in order to minimize scattering losses and three of the mirrors are highly reflective at 1064.5~nm and 849.2~nm, $R > 99.94\%$, while the input coupler has a transmission of $T = 8\%$ at both 1064.5~nm and 849.2~nm. The crystal is placed between the two curved mirrors where the beam waist is the narrowest of the bow-tie cavity. In order to maximize the conversion efficiency, corresponding to an optimization of the Boyd-Kleinman factor \cite{boyd1968parametric}. The beam waist at 852 nm is about 20 \textmu m and the beam waist at 1064 nm is about 22 \textmu m.  This is enabled by using a cavity length of 391 mm and setting the distance between the two curved mirrors to be 45 mm. In order to keep both pump beams on resonance with the cavity, we apply the Pound-Drever-Hall (PDH) locking method by phase modulating the two fundamental pump beams and demodulating the measurements of the cavity reflection \cite{PDH}. The first PDH lock controls the cavity length via an intracavity piezo-actuated mirror to keep the cavity on resonance at 1064.5~nm. The second PDH lock stabilizes the Ti:sapphire laser frequency in order to maintain the double resonance. This stabilization technique locks the Ti:sapphire laser frequency to the more stable fiber laser using the DR cavity as a frequency transfer cavity. At the same time, it ensures a high frequency stability of the 472.4 nm SFG output. The modulation, demodulation, and locking loops are all performed with Red Pitaya boards with a maximum bandwidth of 62.5 MHz \cite{Neuhaus2017}.

\section{Experimental results}

\begin{figure}[h!]
\centering\includegraphics[width=13cm]{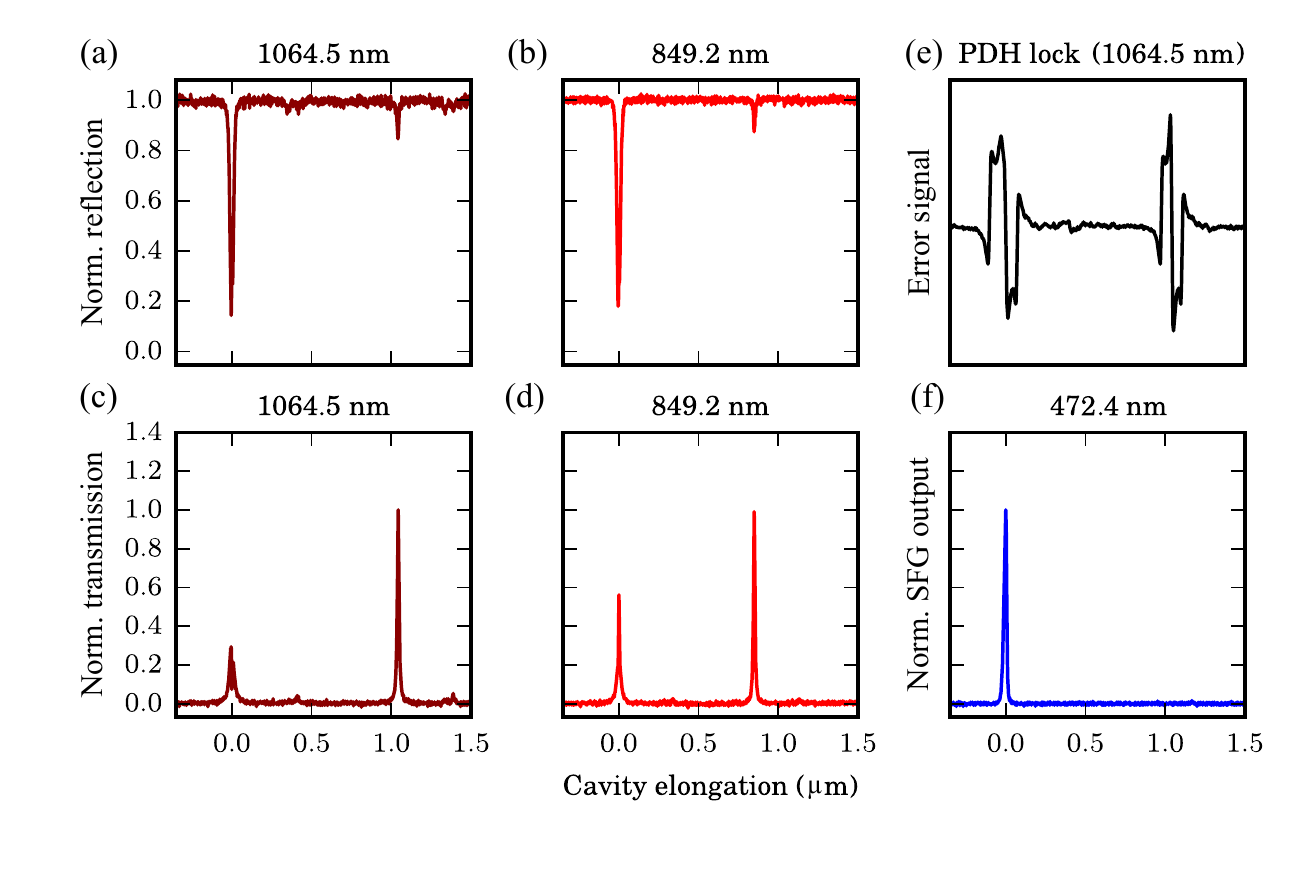}
\caption{Cavity response while scanning the cavity length over two resonances of the TEM$_{00}$ mode, showing the reflection from the cavity input mirror at 1064.5~nm (a) and 849.2~nm (b), transmission through the cavity at 1064.5~nm (c) and 849.2~nm (d), a typical error signal for PDH locking of the cavity length (e), and the intensity of the SFG field at the output of the cavity (f). The traces are normalized to the maximum intensity. At 0 elongation, both pump fields are resonant. The optical pump powers are set such that the pump beams are nearly impedance-matched under depletion by the SFG process. The resonances occurring at elongations matching the pump wavelengths, i.e. 849.2~nm and 1064.5~nm, correspond to single resonant cases in the overcoupled cavity regime.}
\label{Fig3_cavity}
\end{figure}

The cavity transmission and reflection of the pump fields at 849.2~nm and 1064.5~nm are shown in Fig.~\ref{Fig3_cavity}, along with a typical error signal of the PDH locks and the SFG output field intensity at 472.4~nm. When both pump fields are resonant and phase-matching conditions are fulfilled by adjusting the temperature of the crystal, a significant SF power is generated due to the high circulating pump powers. Depletion of the pump fields can be seen as a drop in the intensity of the transmitted field. In addition, the increased intracavity loss alters the coupling of the incident pump fields into the cavity, which can go from overcoupled ($R_{i,\mathrm{in}}<(1-\Delta_i)(1-\Gamma_i)$) to impedance-matched ($R_{i,\mathrm{in}}=(1-\Delta_i)(1-\Gamma_i)$) or undercoupled ($R_{i,\mathrm{in}}>(1-\Delta_i)(1-\Gamma_i)$) depending on the balancing of the pump powers. We infer the mode-matching of the fundamental TEM$_{00}$ modes of the pump beams by measuring the drop of reflected power when the cavity is impedance-matched. We measure a similar mode-matching for the two pump beams of 88.1(0.6)$\%$. The intracavity passive loss were calculated from measurements of the cavity reflection and mode-matching to be $\Delta_{1,2} = 0.3(\pm 0.1)\%$. With a length of 391($\pm 2$)~mm and an input coupler with 8($\pm 1$)$\%$ transmission, the cavity has a free spectral range (FSR) of 767($\pm 4$)~MHz, a finesse of 73($\pm 5$), and a cavity bandwidth of 10.6($\pm 0.7$) MHz.

\begin{figure}[h!]
\centering\includegraphics[width=13cm]{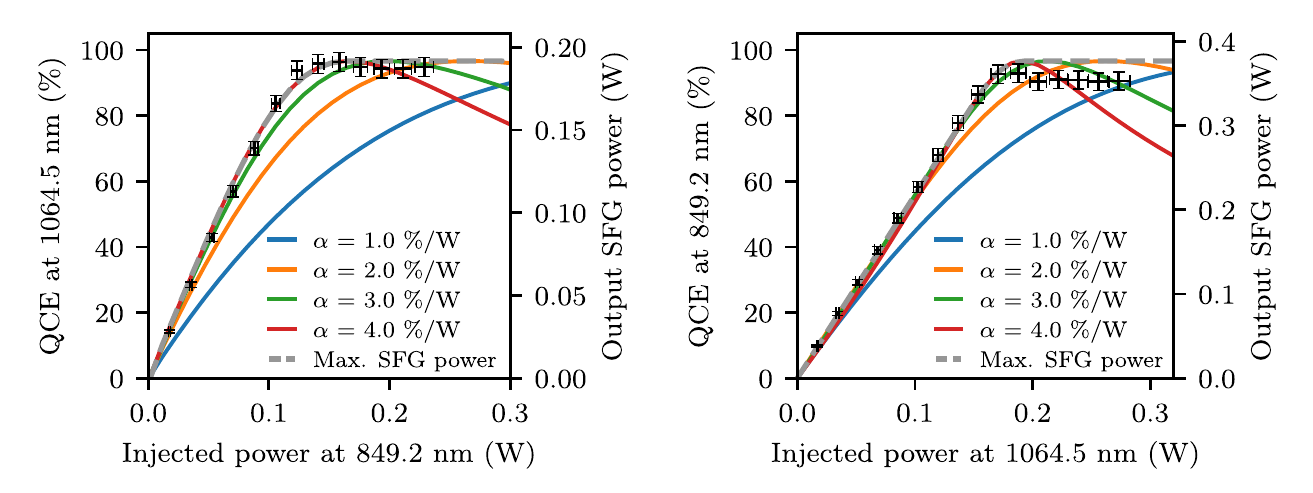}
\caption{QCE and SFG optical power as function of injected pump power at 849.2~nm (a) and 1064.5~nm (b) with the incident pump powers at 1064.5~nm and 849.2~nm fixed to 100 mW and 250 mW, respectively. Symbols denote measured values while solid lines represent simulations with different nonlinear conversion coefficients ($\alpha =$ 0.01, 0.02, 0.03, and 0.04 W$^{-1}$). The dashed line shows the maximum SFG power achievable with nonlinear conversion coefficients up to 0.04 W$^{-1}$. Parameters used in the simulations are $R_{1,\mathrm{in}} = R_{2,\mathrm{in}} = 0.92$, $\Delta _1 = \Delta_2= 0.003$ and 88$\%$ mode-matching.}
\label{Fig4_SFGpower}
\end{figure}

\begin{figure}[t!]
\centering\includegraphics[width=13cm]{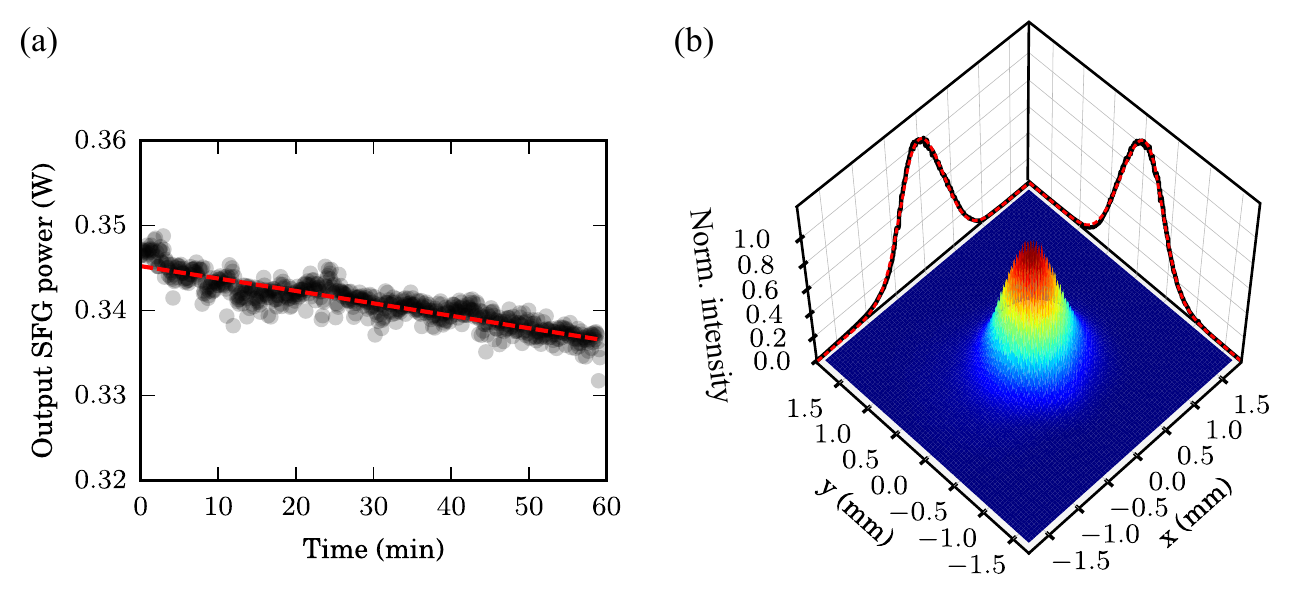}
\caption{(a) Power stability measurement over 1 hour. The red dashed trace is a linear fit to the data. The power stability is measured to be 0.8$\%$ over one hour and the fit indicates a slow decrease of output SFG power of about -2.5 \textmu W/s. The short term stability can be estimated after correction of the slope and reads 0.35$\%$. (b) Beam profile of the SFG output after collimation by a spherical lens with 50 mm focal length. The intensity profile is normalized to the peak intensity. The intensity distribution indicates a single TEM$_{00}$ mode as expected by the cavity design. The cross sections of the beam intensity along the x and y axis (black traces) show very good matching with Gaussian fits (red dashed traces).}\label{Fig5}
\end{figure}

In order to characterize our SFG source and compare with theoretical predictions, we measure the generated blue light power as a function of fundamental pump powers. Figures~\ref{Fig4_SFGpower}(a) and \ref{Fig4_SFGpower}(b) show the generated SF power, while varying the pump power at 849.2~nm and 1064.5~nm and keeping the incident pump powers at 1064.5~nm and 849.2~nm fixed to 100($\pm 3$)~mW and 250($\pm 8$)~mW, respectively. For each data point we optimize the output power by adjusting the crystal temperature. A change of the crystal temperature affects the phase-matching condition of the crystal and therefore changes the nonlinear conversion coefficient of the SFG process. For a resonant SFG system, a maximum of the nonlinear conversion coefficient is not always desirable as it increases intracavity loss and thereby affects the incoupling of the pump beams. This effect is illustrated in Fig.~\ref{Fig4_SFGpower}, where solid lines show SFG powers for fixed nonlinear conversion coefficients, according to theory within the undepleted-pump approximation. In Fig.~\ref{Fig4_SFGpower}, the internal QCE is calculated by considering photons mode-matched to the cavity. Note that we can compare the efficiency of our DR system to single-pass SFG by taking into account the 88$\%$ mode-matching of the pump beams into the cavity, which still provides more than 80$\%$ external conversion efficiency.

We achieve internal QCEs of $\eta_1 = 95(\pm 4)\%$ at 849.2~nm and $\eta_2 =94(4)\%$ at 1064.5~nm with incident powers of 250($\pm 8$) mW at 849.2~nm and 200($\pm 6$) mW at 1064.5~nm, resulting in 375($\pm 11$) mW of output power at 472.4 nm. Comparing these results with our theoretical model we estimate a nonlinear conversion coefficient $\alpha = 0.045(\pm 0.015)~\mathrm{W}^{-1}$ for this measurement. Our measured value of the overall QCE on injected photons $\eta = 95(\pm 3)\%$ is reported on Fig.~\ref{Fig:Theory}(a). Note that in theory 100$\%$ QCE is achievable in a completely lossless systems when the input powers are set such that both pump beams are impedance-matched to the cavity. However losses in the nonlinear optical materials and at cavity mirrors usually sets the upper limit. Absorption and scattering directly reduce the conversion efficiency and limits the maximum value of the cavity finesse, and thus the possible resonant field enhancement. Absorption in the nonlinear crystal leads to heating, thermal lensing, and potentially also a deformation of the spatial mode, which are limiting the conversion efficiency. In our system, the main limitation to the conversion of incident photons is the 88$\%$ mode-matching to the cavity, resulting in an external QCE of 84($\pm 3$)$\%$.

In addition to conversion efficiency, we measured the power stability over time and the output mode profile. Figure~\ref{Fig5}(a) depicts the measured optical power stability of the generated SF light over one hour. The output SFG power shows deviations below 0.8$\%$ over one hour at 341 mW with a sampling interval of one second. It can be seen on Fig.~\ref{Fig5}(a) that the amplitude of the deviations is in part due to a slow decrease of the output SFG power over time, which can be attributed to an increase in intracavity loss due to blue-light-induced infrared absorption and/or grey tracking in the PPKTP crystal \cite{Boulanger1999, Hirohashi2007, Jensen2013, Tjornhammar2015}, and to a slow heating of the crystal that gives rise to effects such as thermal dephasing and thermal lensing \cite{Hansson2000}.However, We have not identified exactly the origin of the slow decline we verified simply that it is not due to instabilities of the pump powers, as they do not follow the same trend over time. The declining rate appeared constant over several hours until the cavity would jump out of lock. The decline was accompanied by increasing loss in the cavity. The loss and the output power could be recovered after moving the crystal, indicating that the declining trend and associated increasing loss are caused by effects in the PPKTP crystal. In order to estimate the short term stability when the slow decrease of SFG power is negligible, we apply a linear fit with a slope of -2.5 \textmu W/s and find a standard deviation from the fit line of 0.35$\%$. The high stability of our system results from the robust mechanical design of the cavity and the implementation of efficient locking techniques.

Diffraction-limited blue lasers are attractive due to their tight focusing capabilities within the spectral response of common silicon detectors. We therefore investigate the beam quality of our SF output. Figure~\ref{Fig5}(b) shows the beam profile of the SFG output after collimation by a spherical lens with 50 mm focal length. The SFG output is in a TEM$_{00}$ mode with a two-dimensional Gaussian intensity profile. A measurement of the M$^2$ value by focusing the output SF beam gives a value below 1.05, thereby demonstrating the good quality of our SF beam for applications such as microscopy.

\section{Conclusion}

We have demonstrated highly efficient DR-SFG in the blue wavelength range. With an internal QCE of 95(3)$\%$ of the mode-matched photons, this result constitutes, to the best of our knowledge, a new record of QCE for SFG. The external QCE of 84(3)$\%$ of the incident photons is mainly limited by the 88$\%$ mode-matching to the cavity and can be improved by mode-cleaning of the pump laser beams. The DR-SFG system enables high optical-to-optical conversion efficiency even at moderately low pump powers making it interesting for quantum frequency conversion with a potential very high conversion fidelity. Optical power up to 375 mW at 472.4 nm are produced in a TEM$_{00}$ by coupling 88$\%$ of 250~mW at 849.2~nm power and 200~mW at 1064.5~nm into the bow-tie cavity. Power stability measurements show an optical power deviation of 0.8$\%$ over one hour, which consists mostly of a steady decline of -2.5 \textmu W/s imputable to slow thermalization of the system and probably also to an increasing loss by grey tracking mechanism. The stability of the system is ensured by a robust mechanical design and efficient locking techniques. Measurements of the beam profile of the SFG light shows a diffraction limited TEM$_{00}$ mode with M$^2$ < 1.05. The high spatial quality of the SFG output is readily provided by the DR cavity design and makes our SFG system useful as a light source for various types of microscopes, high spatial resolution scatterometry and dark-field wafer inspection. The wavelength tuning of the SFG output is limited by the phase-matching of the nonlinear crystal to about 5 nm. However, this can be expanded to cover the entire blue spectrum (420 nm to 510 nm) by proper choice of second-order nonlinear crystals. Our SFG system can be adapted for applications with optical atomic clocks that use wavelengths of 457 nm and 461 nm for laser-cooling of Mg and Sr atoms, respectively \cite{Galbacs2006}, and for life science and biomedical research, where various fluorophores are known to excite close to the 488 nm wavelength. Our SFG source can also find application for pumping optical parametric oscillators for the generation of tunable non-classical light in the near-infrared region \cite{Khalili2018}.

\section*{Funding}
The Danish Agency for Institutions and Educational Grants. The Eureka turbo project "Quantum-Gravity Wave Detection" (EUROS E11677 T-Q-GWD). Coordena\c{c}\~ao de Aperfei\c{c}oamento de Pessoal de N\'ivel Superior (CAPES) process (88881.188944/2018-01). Conselho Nacional de Desenvolvimento Cient\'ifico e Tecnol\'ogico (CNPq) process (161702/2015-5). The European Research Council (ERC) under the European Union's Horizon 2020 research and innovation programme (grant agreement 787520)

\section*{Acknowledgments}
We acknowledge Jan Thomsen, NBI, for lending us his Ti:sapphire laser.

\section*{Disclosures}
The authors declare no conflicts of interest.




\end{document}